\documentclass{PoS}
\usepackage{color,natbib}
\title{PROBING THE EXTRAGALACTIC COSMIC RAYS ORIGIN WITH GAMMA-RAY AND NEUTRINO BACKGROUNDS}

\ShortTitle{Multimessenger approach to UHECR origin}

\author{\speaker{Noemie Globus}\thanks{\bf\color{magenta} Oral presentation CRI184 scheduled on Tue., July 18.}\\
        Racah Institute of Physics, The Hebrew University, 91904 Jerusalem, Israel\\
        E-mail: \email{noemie.globus@mail.huji.ac.il}}

\author{Denis Allard\\
        Laboratoire Astroparticule et Cosmologie, Universit\'e Paris Diderot/CNRS, 10 rue A. Domon et L. Duquet, F-75205 Paris Cedex 13, France\\
        E-mail: \email{allard@apc.in2p3.fr}}
        
\author{Etienne Parizot\\
        Laboratoire Astroparticule et Cosmologie, Universit\'e Paris Diderot/CNRS, 10 rue A. Domon et L. Duquet, F-75205 Paris Cedex 13, France\\
        E-mail: \email{parizot@apc.in2p3.fr}}
        
\author{Tsvi Piran\\
        Racah Institute of Physics, The Hebrew University, 91904 Jerusalem, Israel\\
        E-mail: \email{tsvi.piran@mail.huji.ac.il}}                

\abstract{ GeV-TeV gamma-ray and PeV-EeV neutrino backgrounds provide a unique window on the nature of the ultra-high-energy cosmic-rays (UHECRs).
We discuss the implications of the recent Fermi-LAT data regarding the extragalactic gamma-ray background (EGB) and related estimates of the contribution of point sources as well as IceCube neutrino data on the origin of the UHECRs. We calculate the diffuse flux of cosmogenic $\gamma$-rays and neutrinos produced during the UHECRs propagation and derive constraints  on the possible cosmological evolution of UHECR sources.  In particular, we show that the mixed-composition scenario which is in agreement with both (i) Auger measurements of the energy spectrum and composition up to the highest energies and (ii) the ankle-like feature in the light component detected by KASCADE-Grande, is compatible with both the Fermi-LAT measurements and with current IceCube limits. 
}

\FullConference{35th International Cosmic Ray Conference --- ICRC2017\\
		10--20 July, 2017\\
		Bexco, Busan, Korea}

\begin{document}

\section{Introduction}
\label{sec:intro}

The interaction of ultra-high-energy cosmic-rays (UHECRs) with the photon backgrounds during their propagation in intergalactic space produces 
cosmogenic $\gamma$-ray photons 
and neutrinos ($\nu$s).
The flux of these secondary messengers is highly sensitive to the spectral shape, maximal energy, composition and cosmological evolution of the UHECR sources, and therefore, one can derive  important constraints on the UHECR origin from a multi-messenger approach that takes these into account
\citep[for $\gamma$-rays]{Protheroe96,CoppAha97,
Ahlers11,2011A&A...535A..66D,Bere16, Supanitsky16,2016ApJ...822...56G}; \citep[e.g.][for $\nu$s]{Stecker79, 
Engel01, Seckel05, Allard06, 
Anchordoqui07, Ahlers09, Kotera10}. 


The recent Fermi-LAT 
data \citep{2015ApJ...799...86A}, together with 
statistics of the photon counts in the skymap pixels
\citep[e.g.][and references therein]{Malyshev11} have enabled different authors \citep[][hereafter A16 and Z16]{Ack2016,Zechlin16} to estimate the flux contributed by point sources (PS) well below the Fermi-LAT detection limits. 
These studies show that resolved and unresolved PS account for the majority of the EGB. 
Since a $\gamma$-ray background due to  extragalactic cosmic rays (EGCRs) is unavoidable, it is crucial to verify that the proposed UHECR source models do not violate the existing constraints.

Recent measurements by the Pierre Auger Observatory (Auger) 
indicate that the composition of UHECRs is mixed (predominantly light) at the ankle of the cosmic-ray spectrum, and it gets progressively heavier as the energy increases 
 \citep{Aab14b
 }. This composition trend can be interpreted as the signature of a low maximal energy-per-unit-charge ($E_{\rm max}/Z\lesssim  10^{19}$~eV) 
of the nuclei accelerated at the dominant sources of UHECRs. 
Below $10^{18}$ eV, the KASCADE-Grande experiment 
reported an ankle-like feature in the energy spectrum of light (proton-helium) elements with a break at $\sim 10^{17}$~eV \citep{Apel13,Bertaina15}.
This 
``light ankle'' can be naturally understood as the emergence of a light EGCR component, 
taking over the steeper Galactic cosmic-ray (GCR) component. 

We discuss here the viability of a class of mixed-composition models in which the KASCADE-Grande and Auger data are understood in terms of a transition between a GCR component and a single EGCR component with a soft proton spectrum and low $E_{\rm max}$. 
This soft proton component, responsible for the KASCADE-Grande light ankle, would be the dominant contributor to the cosmogenic $\gamma$-ray flux.
This model was shown to be compatible with the spectrum and composition data at all energies \citep[][hereafter G15a, G15b]{Globus2015a, Globus2015b}, and it is consistent with the anisotropy constraints on galactic protons \citep{Tinyakov16}.


\section{\label{sec:model}Source model and propagated cosmic-ray spectrum}

Any phenomenological EGCR model that account for the data needs a very hard spectrum at the sources, to reproduce the evolution of the composition above the ankle observed by Auger, and a softer proton component, to account for the light ankle seen by KASCADE-Grande.

  \begin{figure}[ht!]
 \centering
  \includegraphics[width=0.46\linewidth]{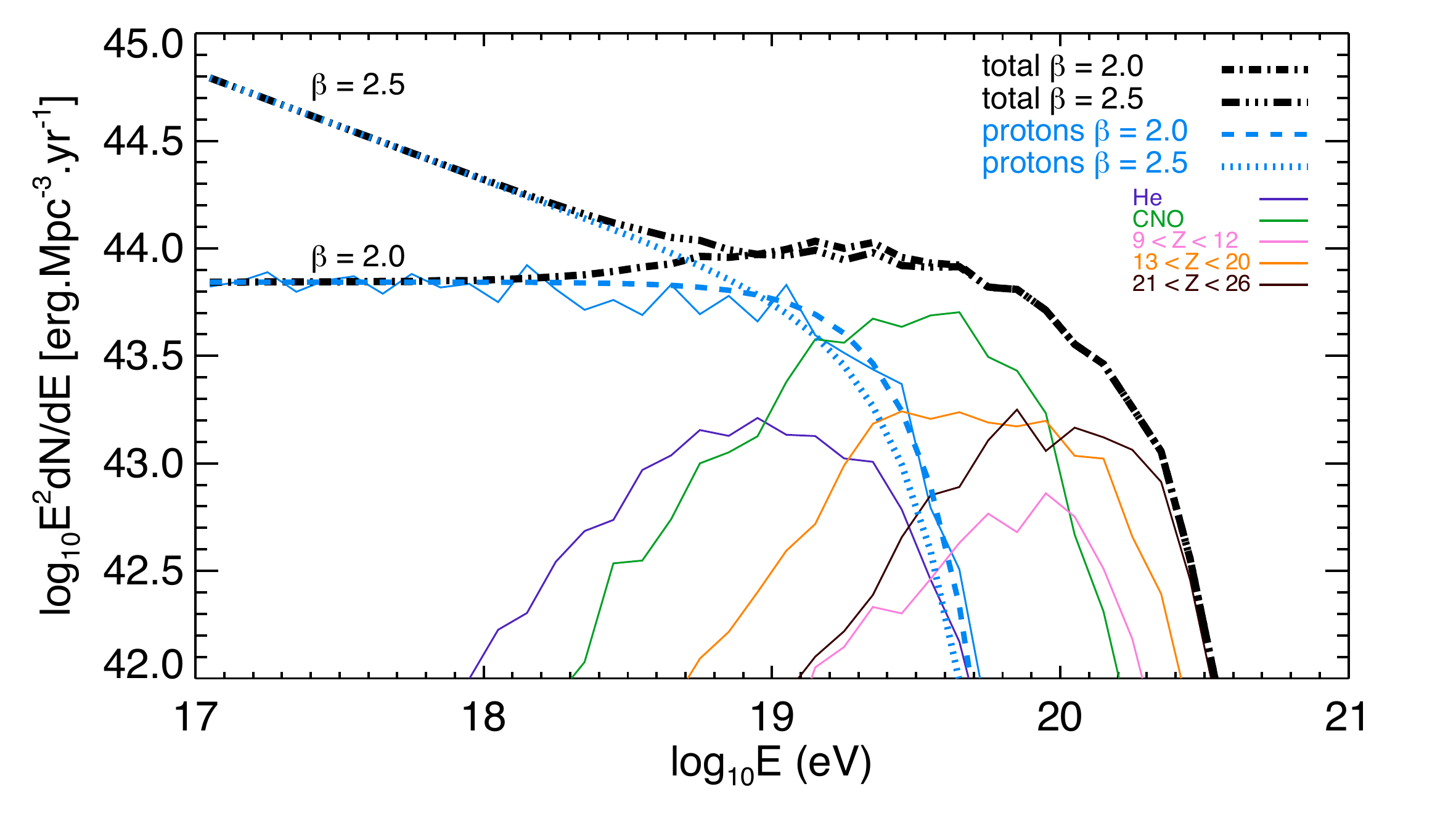}
  \includegraphics[width=0.46\linewidth]{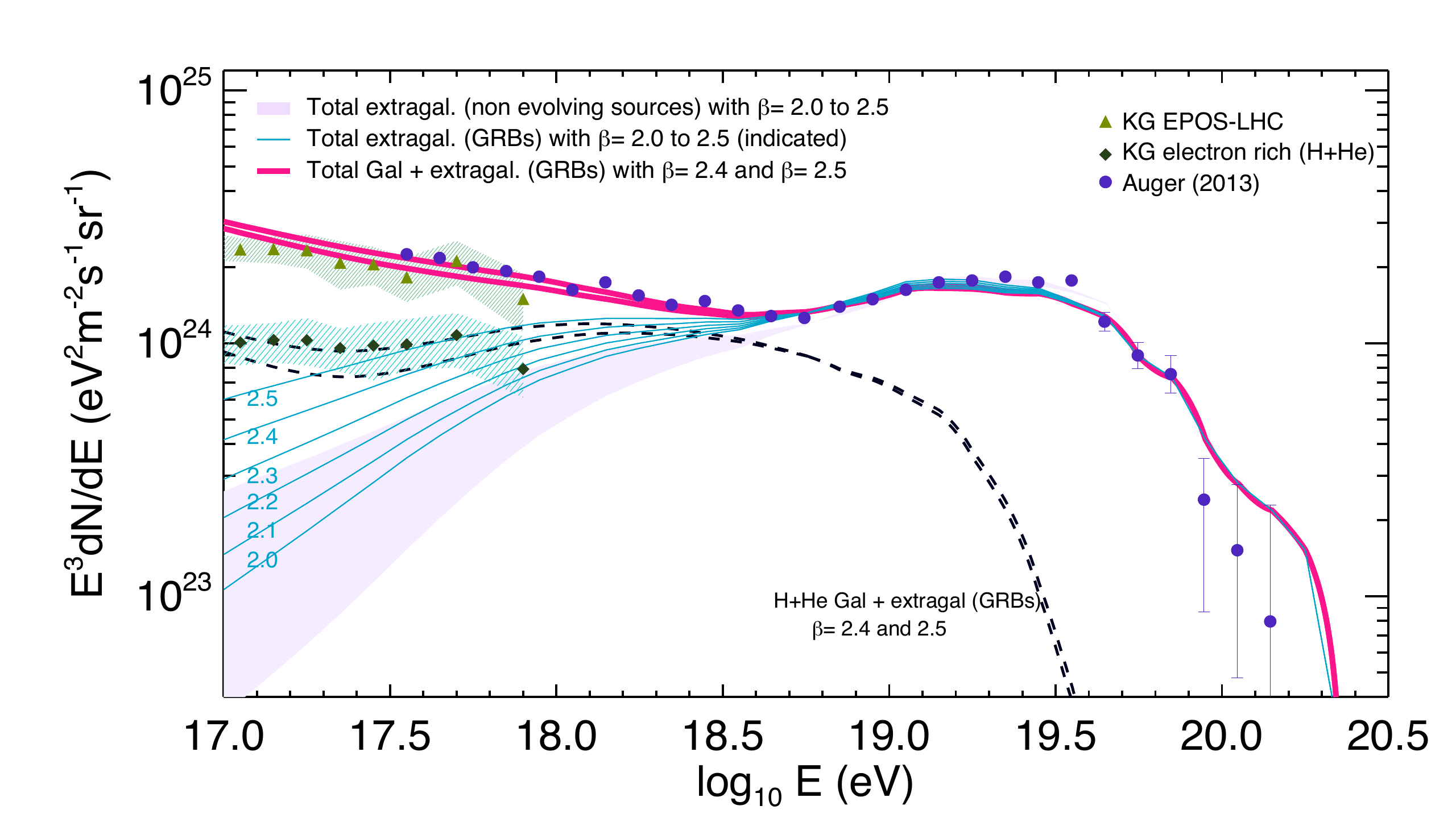}
 \caption{Left panel: UHECR injection spectrum for the various nuclei, as obtained in G15a, with the fit of the proton component with spectral index $\beta = 2.0$ (dashed blue), and with its modified shape in the case of $\beta=2.5$. Right panel: Propagated UHECR spectra  for $2.0\leq\beta\leq2.5$, compared to KASCADE-Grande and Auger data, for a GRB-like cosmological evolution (blue lines) or non-evolving sources (violet shaded area). The total (GCR+EGCR) light component  is compared to that deduced from KASCADE-Grande data (using the EPOS-LHC \citep{EPOS1} hadronic model), for GRB-like evolution with $\beta=2.4$ and 2.5 (dashed lines).}
\label{fig:modelc} 
\end{figure}

The effective spectrum from the mixed composition model (G15b) is displayed in the left panel of Fig.~\ref{fig:modelc}.  The much softer source spectrum for the nucleons is due to the free escape of neutrons produced by the photo-disintegration of nuclei at the source, and its exact shape depends on various physical parameters (see G15a for details).

 Since 
the extragalactic protons around $10^{17}$~eV contribute significantly to 
the expected cosmogenic $\gamma$-ray flux in the Fermi energy range, 
we explore, for the sake of generality, (i) different slopes for the  proton component $\beta$ 
(as could result from different physical parameters describing the sources) 
while keeping the same maximal rigidity  and 
spectral shape for heavier nuclei; 
 (ii) different cosmological evolutions, assuming an average source power proportional to $(1+z)^{\alpha}$ up to a redshift $z_{\max}$.
We  consider a range of spectral indices  $2.0\leq\beta\leq2.5$.  The two proton spectra with the  extreme values of $\beta$  are represented  by thick dashed and dotted blue lines, respectively. The implied range of UHECR emissivities above $10^{17}$~eV is  $L_{\rm CR}^{17}\sim[5.7-14]\cdot10^{44}\,\mathrm{erg}\,\,\mathrm{Mpc}^{-3}\,\, \mathrm{yr}^{-1}$.


The right panel of Fig.~\ref{fig:modelc} depicts the \textit{propagated} UHECR spectra for $2.0\leq\beta\leq2.5$,
for EGCR sources evolving as GRBs \citep[][blue lines]{2010MNRAS.406.1944W} and for non evolving sources (violet shaded area). 
In the case of a GRB-like or SFR-like cosmological evolutions, 
proton spectral indices $\beta \simeq 2.4-2.5$ provide a good fit to the KASCADE-Grande data when summing the light EGCR component with the GCR light component obtained in G15b (dashed line in the right panel of Fig.~\ref{fig:modelc}). 
Softer proton indices are required in the case of a non-evolving scenario.


 \begin{figure} [t]
 \centering
  \includegraphics[width=0.42\linewidth]{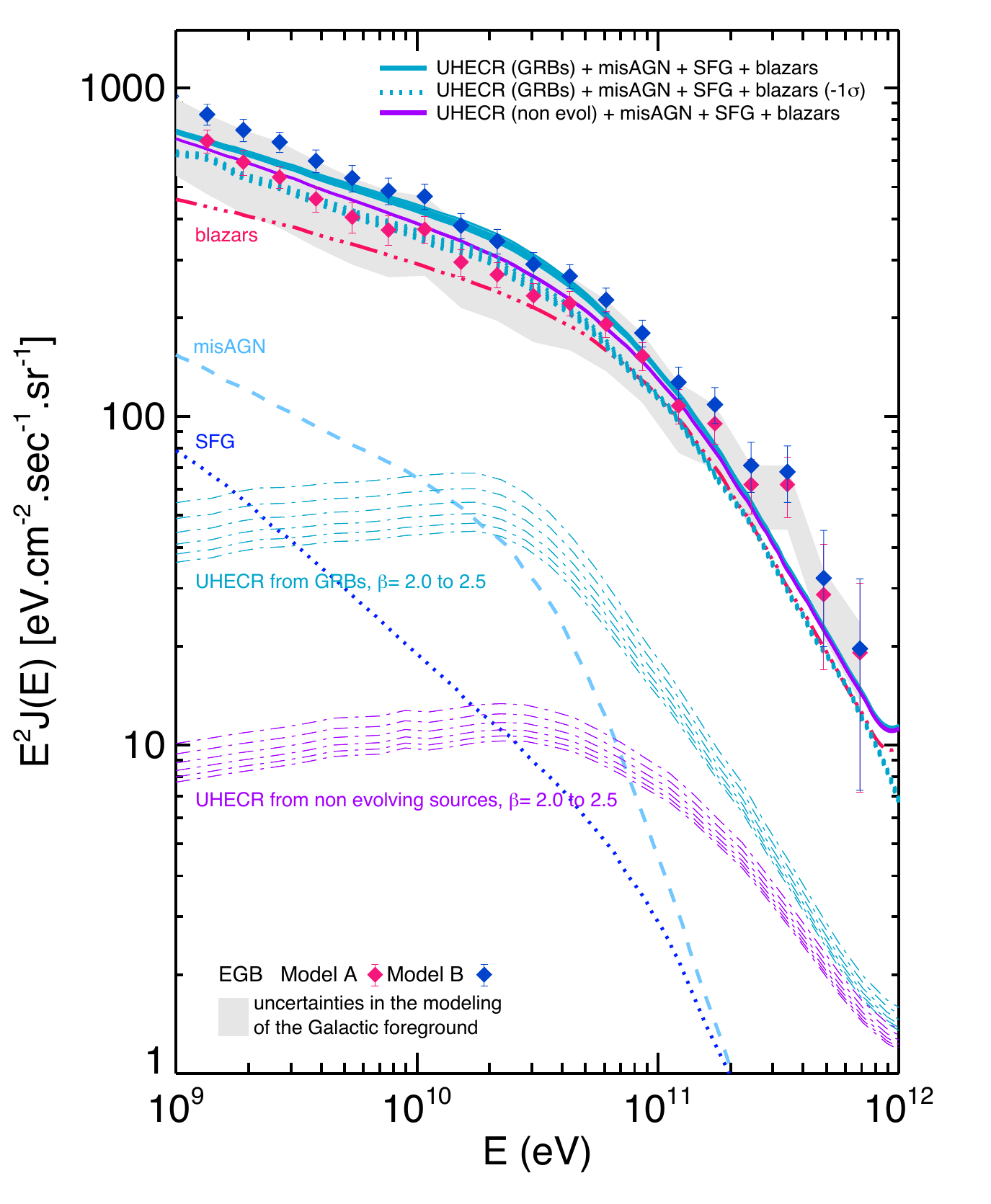}
  \includegraphics[width=0.42\linewidth]{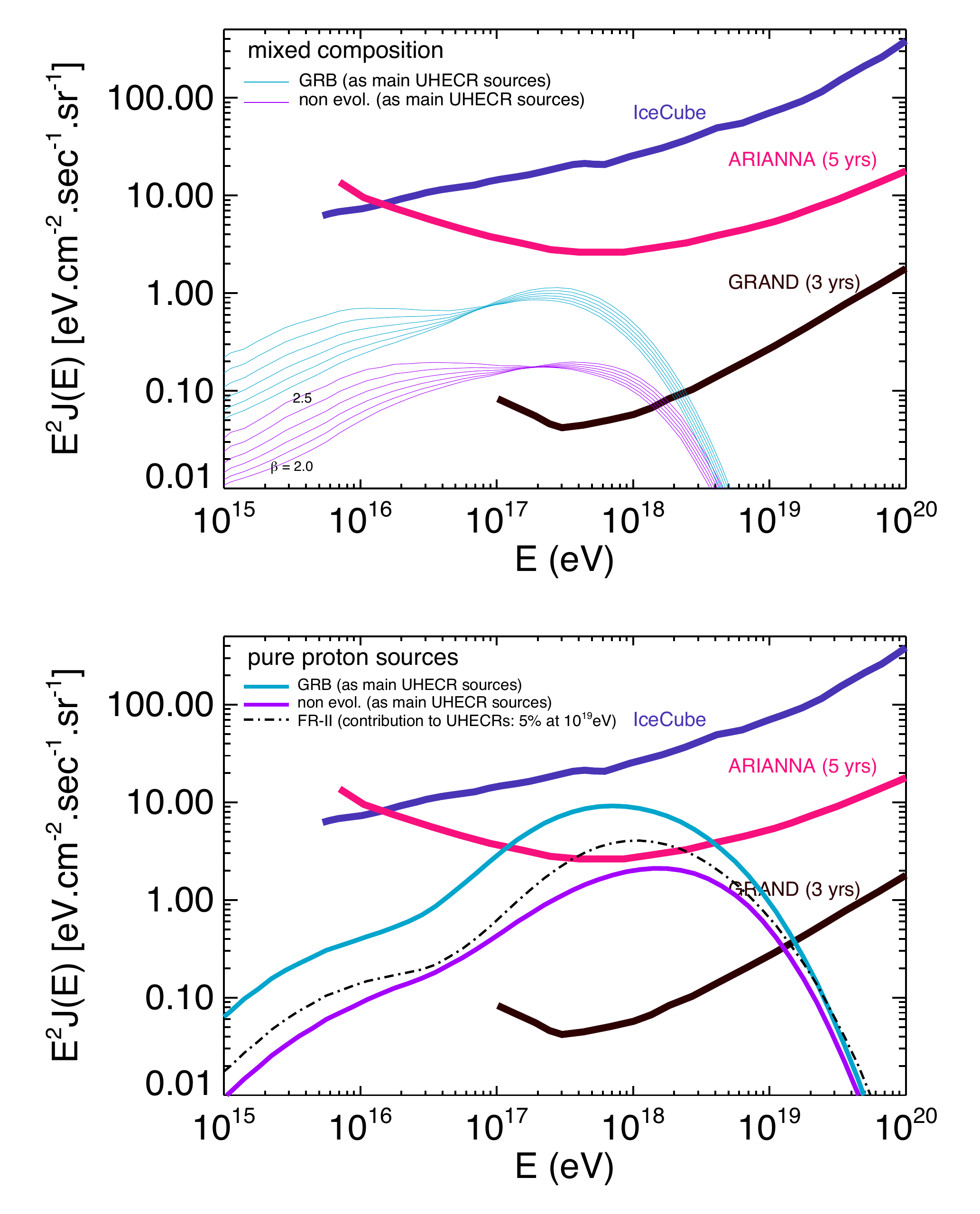}%
 \caption{
Left panel: $\gamma$-ray fluxes from EGCRs (dashed-dotted lines), for GRB-like evolution (blue) and non-evolving (violet) sources, as computed with our mixed-composition model and spectral indices of the soft proton component $2.0\leq\beta\leq2.5$.
Also represented the $\gamma$-ray fluxes from SFG, MisAGN and blazar sources (see labels) as modelled by \cite{Ack2012,Inoue11,Ajello15} respectively. The corresponding sum of UHECR, SFG, misAGN and blazar components is represented by thick solid lines  \citep[or with a dotted line when 1-$\sigma$ lower bound are adopted for the SFG+misAGN+blazar model, see][]{Ajello15}, and compared to the EGB estimated from Fermi-LAT data, for both foreground models A and B.  
Right upper panel: cosmogenic $\nu$ fluxes associated with our mixed-composition scenarios in the case of GRB-like evolution (blue) and non-evolving (violet) sources, compared with the current IceCube sensitivity \citep{Aartsen16} and the expected sensitivities of ARIANNA (5 years, 50 MHz option, \cite{Hallgren16}) and GRAND (3 years, \cite{Martineau15}). Right lower panel: same, i) for 100\% proton scenarios compatible with the Fermi constraints (plain lines, same colour code), and ii) for a sub-dominant proton component (contributing 5\% of the UHECRs at 10~EeV) evolving as FR-II galaxies (dashed-dotted line) \citep{Wall05}.
 }
 \label{fig:EGB}
 \end{figure}

\section{Gamma-ray and neutrinos counterparts}

\label{sec:gamma}
The interactions of the propagating  EGCRs leads to the production of cosmogenic $\gamma$-rays in the GeV-TeV range, and $\nu$s in the PeV-EeV range, through the development of electromagnetic cascades.
 The Monte-Carlo procedure used to calculate the cosmic-ray, $\nu$ and $\gamma$-ray spectra is presented in \citet{2011A&A...535A..66D}.

The cosmogenic $\gamma$-rays spectra corresponding to the UHECR spectra of Fig.~\ref{fig:modelc}, 
are shown  in Fig.~\ref{fig:EGB} for a mixed-composition model with proton spectral indices $2.0\leq\beta\leq2.5$, for sources 
 with no cosmological evolution (violet lines) and with a GRB-like evolution (in blue). 
These  $\gamma$-ray fluxes represent only a small contribution to the total EGB, which is reproduced from \citet{2015ApJ...799...86A} for two different models of the Galactic $\gamma$-ray foreground, referred to as model A and model B by the authors. 
These two models roughly differ by $\sim 20-30\%$, which can be seen as a rough estimate of their systematics in the subtraction process.

To determine whether a given EGCR source model  is  compatible with the $\gamma$-ray data, we need to take into account other known contributions to the EGB. 
The contribution of star-forming galaxies (SFG) and misaligned active galactic nuclei (misAGN), based on the models by \citet{Inoue11} and \citet{Ack2012}, are shown in the left panel of Fig.~\ref{fig:EGB} (omitting the uncertainty bands for clarity).  Also shown 
 is the $\gamma$-ray spectrum arising from blazars, adapted from \citet{Ajello15}, which appears in good agreement with the PS contribution estimated by  A16 and Z16 over the whole energy range.
We find that, 
for  the SFR and weaker evolving scenarios, the sum of all components (UHECR, misAGN, SFG and blazars) never exceeds the total EGB, in the case of model B. 
In the case of model A, the sum is above the EGB. However, it falls below it if one adopts the $1\sigma$ lower bound on the misAGN+SFG+blazars contribution (see \citet{Globus2017} for the estimates of the fluxes of the different contributions in the same energy bands as in A16 and Z16).

The right panel of Fig.~\ref{fig:EGB} shows the resulting $\nu$ spectra for different EGCR models, together with the sensitivity of current and planned experiments.
The mixed-composition models predict $\nu$ fluxes too low to be detected by IceCube \citep{Aartsen16} or ARIANNA \citep{Hallgren16}, even in the case of a GRB-like cosmological evolution.
They would require a sensitivity such as that expected for the GRAND observatory \citep{Martineau15} or CHANT satellite concept \citep{Neronov16}. 
Pure proton scenarios can be seen on Fig.~\ref{fig:EGB} to yield detectable fluxes, while still being allowed by the current IceCube limits and Fermi-LAT data. (For these calculations, we assumed a pure proton $E^{-2}$ spectrum with an exponential cutoff at $E_{\rm max} = 60$~EeV, which is known to reproduce reasonably well the Auger spectrum above the ankle). 
It is interesting to note that the $\nu$s that could be produced by a hypothetical subdominant EGCR proton sources, with large enough $E_{\rm max}$ and cosmological evolution, would  contribute a detectable  $\nu$ flux around $10^{18}$~eV, thus making EeV $\nu$s a powerful probe for revealing the existence of trans-GZK proton accelerators, even if they do not dominate the observed UHECR flux.

\begin{figure}[!h]
\centering
\includegraphics[width=0.65\linewidth]{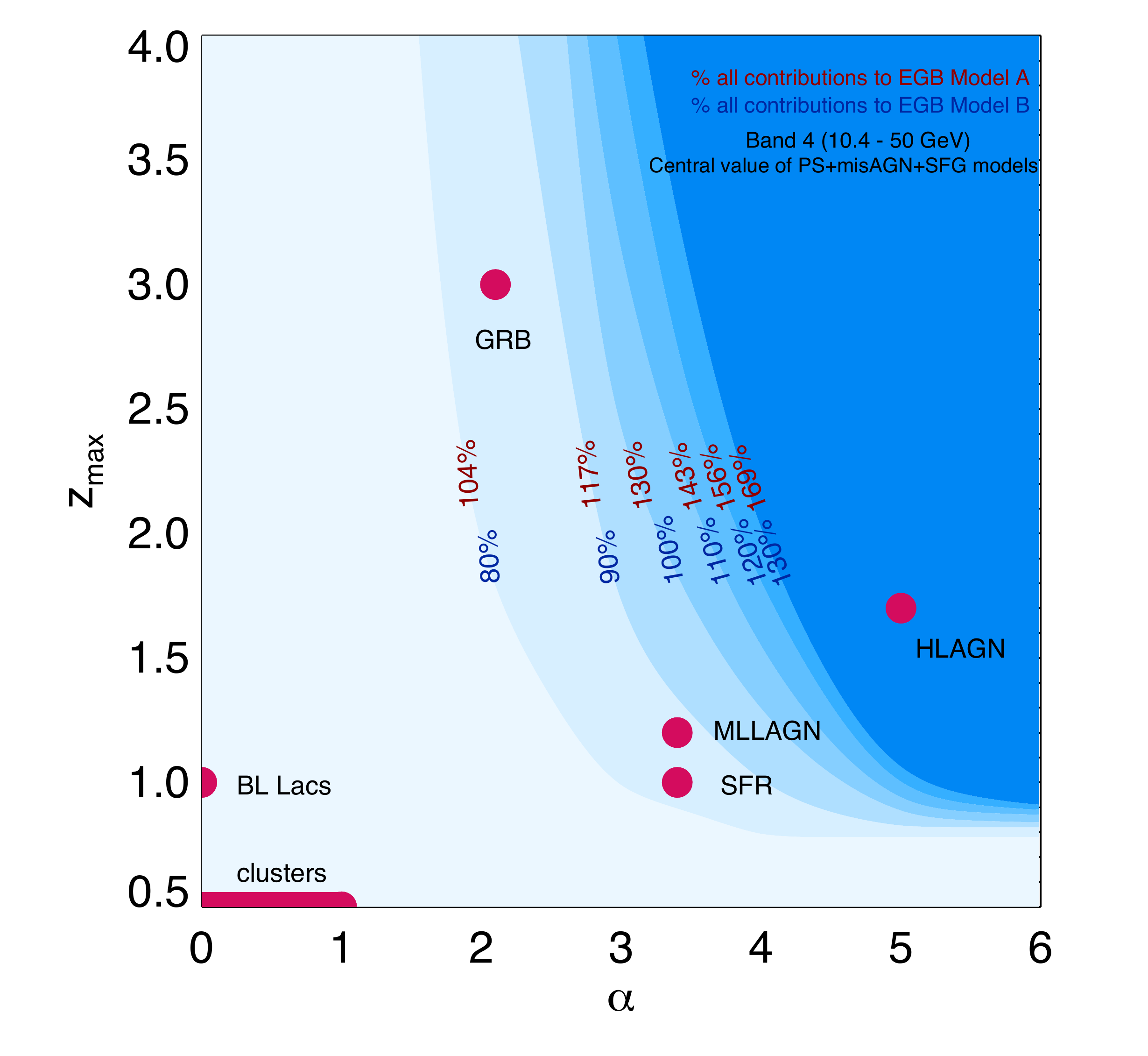}
\caption{Fermi-LAT constraints on EGCR source evolution in the case of our mixed-composition scenario and proton index $\beta=2.5$. 
The different colors show the percentage of the sum of all components (UHECR+PS+misAGN+SFG) to the EGB (Models A and B) in the 10.4--50~GeV energy band, in the ($\alpha$,~$z_{\max}$) parameter space, where $z_{\max}$ is the maximum redshift up to which sources experience a cosmological evolution in $(1+z)^{\alpha}$. Some possible EGCR sources \citep[see e.g.][for the references to the cosmological evolutions]{2016ApJ...822...56G} are shown. 
GRB: gamma-ray bursts. SFR: star-formation rate. MLLAGN: Medium-low-luminosity AGNs. MHLAGN: Medium-High-Luminosity AGNs. HLAGN: High Luminosity AGNs.
}
\label{fig:summary} 
\end{figure}

\section{Conclusions}

The UHECR model considered in G15b  
gives a coherent picture of the GCR-to-EGCR transition,
and  appears to be compatible with the Fermi-LAT measurements and the estimates of the PS contributions by A16 and Z16. 
The mixed-composition model appear to be less constrained by the Fermi-LAT 
 than the electron-positron dip (pure-proton) scenario \citep{Bere16, Supanitsky16, 2016ApJ...822...56G} that rules out SFR-like and stronger cosmological evolutions (see also \cite{Heinze16} for more radical conclusions on the dip model).
 Our results are summarized in Fig.~\ref{fig:summary}, that shows the allowed parameter space of different evolutionary scenarios for the mixed composition model. This estimate is based on the summed contribution of all components in the 10.4-50 GeV band, where the contribution from UHECRs is the largest. Only very strong evolutions, e.g. similar to the very luminous AGNs, are excluded by the current observations.
For the evolutionary models allowed by Fermi, the $\nu$s fluxes above $10^{17}$~eV associated with the mixed-composition scenario are well below the current  IceCube limits. These fluxes are 
within the reach only of the most sensitive planed $\nu$ observatories.

Finally, we note that while the PS contributions are now understood to dominate the extragalactic $\gamma$-ray fluxes in the GeV-TeV range, the uncertainties on the different contributions \citep[notably for  sources other than blazars, see e.g][]{DiMauro13,Lacki14,Tamborra14} as well as on the Galactic foreground are still too large to efficiently constrain the cosmological evolution of UHECR sources. Since the $\gamma$-ray fluxes  associated  with mixed-composition UHECRs never exceed $\sim 20$\% of the EGB  (at least for source evolutions not significantly larger than SFR, see Table~2 in \citet{Globus2017}),  the EGB and its other contributions should be  { determined } to this level of precision in order to estimate whether a UHECR mixed-composition model is  excluded. 
Moreover, the Fermi-LAT estimates of the Galactic foreground are based on the GALPROP framework \citep{Strong00}. These calculations rely on several simplifying assumptions in particular in the description of the Galactic cosmic-ray source distribution or the magnetic halo, as well as on several {\it ad-hoc} parameters that are  tuned to reproduce cosmic-ray data. Alternative models 
\citep[e.g.][and references therein]{Nava17}
have been shown to fairly account for the primary-to-secondary ratios as well as some puzzling features in the observed $\gamma$-ray Galactic signal. 
These models have a smaller halo extension and would probably result in a lower Galactic foreground, leaving more room for EGCR contributions.

\begin{acknowledgments}
NG and TP acknowledges the I-CORE Program of the Planning and Budgeting Committee and The Israel Science Foundation (grant 1829/12), the advanced ERC grant TReX, and the Lady Davis foundation. 
\end{acknowledgments}

\end{document}